\documentclass[12pt]{article}

\usepackage{fancybox}
\usepackage{cite}
\usepackage{float}
\usepackage{amsfonts}
\usepackage{amsmath}
\usepackage{amsbsy}
\usepackage{graphicx}
\usepackage{amssymb}
\usepackage{amsthm}
\usepackage{bm}
\usepackage{epsfig}
\usepackage{latexsym}
\usepackage{pdflscape}
\usepackage{color}
\usepackage{here}
\usepackage{graphicx}
\numberwithin{equation}{section}

\allowdisplaybreaks

\def\x{{\bm x}}
\def\y{{\bm y}}
\def\z{{\bm z}}
\def\llangle{\langle\!\langle}
\def\rrangle{\rangle\!\rangle}
\DeclareMathOperator{\Det}{Det}

\begin{document}

\renewcommand{\thefootnote}{\fnsymbol{footnote}}

\begin{titlepage}
\begin{flushright}
{\footnotesize OCU-PHYS 443}
\end{flushright}
\bigskip
\begin{center}
{\LARGE\bf Giambelli Identity\\[8pt]
in Super Chern-Simons Matrix Model}\\
\bigskip\bigskip
{\large 
Satsuki Matsuno\footnote{smatsuno@sci.osaka-cu.ac.jp}
\quad and \quad
Sanefumi Moriyama\footnote{moriyama@sci.osaka-cu.ac.jp}
}\\
\bigskip
{\it Department of Physics, Graduate School of Science,
Osaka City University\\
3-3-138 Sugimoto, Sumiyoshi, Osaka 558-8585, Japan}
\end{center}

\bigskip

\begin{abstract}
A classical identity due to Giambelli in representation theory states that the character in any representation is expressed as a determinant whose components are characters in the hook representation constructed from all the combinations of the arm and leg lengths of the original representation.
Previously it was shown that the identity persists in taking, for each character, the matrix integration in the super Chern-Simons matrix model in the grand canonical ensemble.
We prove here that this Giambelli compatibility still holds in the deformation of the fractional-brane background.
\end{abstract}
\centering\includegraphics[scale=0.6]{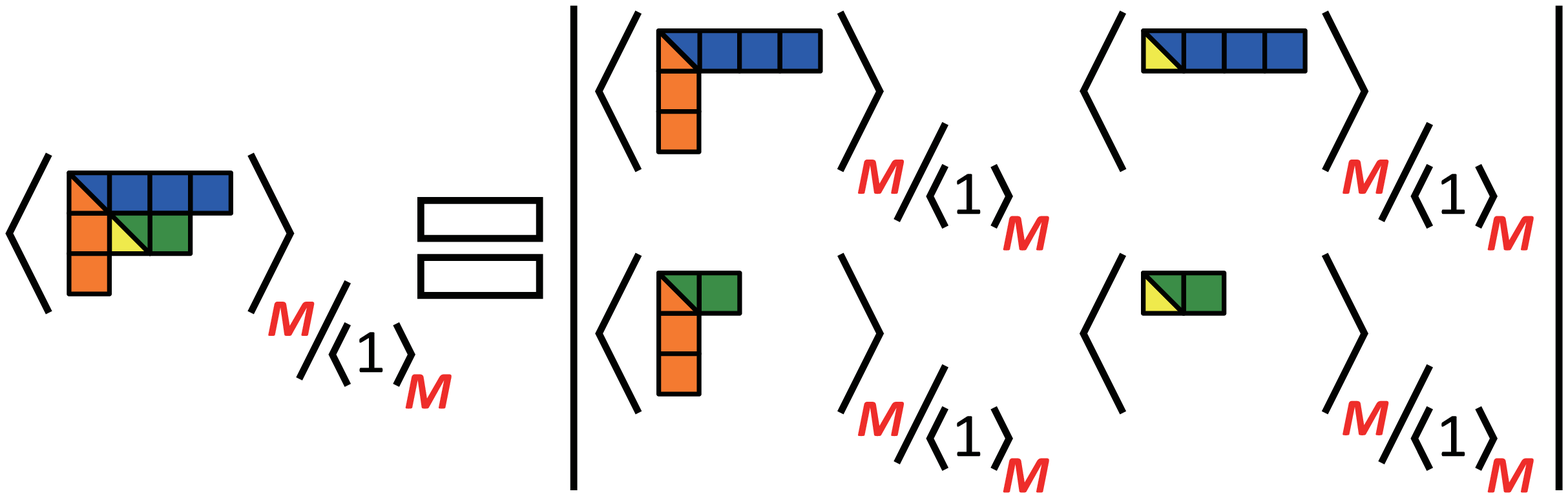}
\end{titlepage}

\renewcommand{\thefootnote}{\arabic{footnote}}
\setcounter{footnote}{0}

\section{Introduction}\label{intro}

Young diagram has a deep relation with the free fermion system.
This can be seen clearly by transforming the Young diagram into a Maya diagram, a line of black and white circles, which is depicted by the following rule.
Namely, we trace the boundary of the Young diagram and draw a black circle when going vertically and a white circle when going horizontally.
See figure \ref{maya} for examples.
If we regard the black circle as a state occupied by a fermion and the white circle as an unoccupied state, this matches exactly with the picture of the fermion excitations.
Here the trivial diagram corresponds to the vacuum state where there are no excitations from the Dirac sea, while a general diagram corresponds to a general excited state.

One main characteristic of these free fermions is the Giambelli identity, which is reminiscent of the Slater determinant for a system of multiple free fermions.
The Giambelli identity states that a character in any representation is expressed as a determinant whose components are characters in the hook representation constructed from all the combinations of the arm lengths and the leg lengths of the original representation.
Due to the universal property of the free fermions, we expect that the Giambelli identity is ubiquitous.
In fact, the Giambelli identity \cite{gi} originally found for the Schur polynomial, the U$(N)$ character, is also valid for the super Schur polynomial, the U$(N_1|N_2)$ character \cite{md}.
Sometimes a quantity defined in a much more complicated way shares this property.
When the Giambelli identity holds for a quantity, we call the quantity Giambelli compatible \cite{BOS}.
For example, as we shall explain later, the one-point function of the half-BPS Wilson loop in the ABJM theory in the grand canonical ensemble is Giambelli compatible.
In this work, we find that the Giambelli compatibility persists in a shift of ``background'' parametrized by an integer $M$.

However, the proof of the Giambelli compatibility is in general not easy.
In this work, we propose an easier criterion for the Giambelli compatibility shifted by an integral parameter $M$.
We shall see that, as long as a quantity is given in another determinantal expression (which is a simple generalization of the Giambelli compatibility for $M=0$ and seems simpler to prove in many cases), the Giambelli compatibility holds for general $M$.

In the next subsection, we start with some preparations to state our mathematical criterion.
Then, in the subsequent subsection, we shall explain that this criterion is easier to prove in the ABJM theory and maybe others.
In section \ref{pf} we present a proof for our mathematical criterion.
Finally we conclude with some discussions in section \ref{conclusion}.

\subsection{Mathematical formulation}\label{math}

\begin{figure}[!ht]
\centering\includegraphics[scale=0.6,angle=-90]{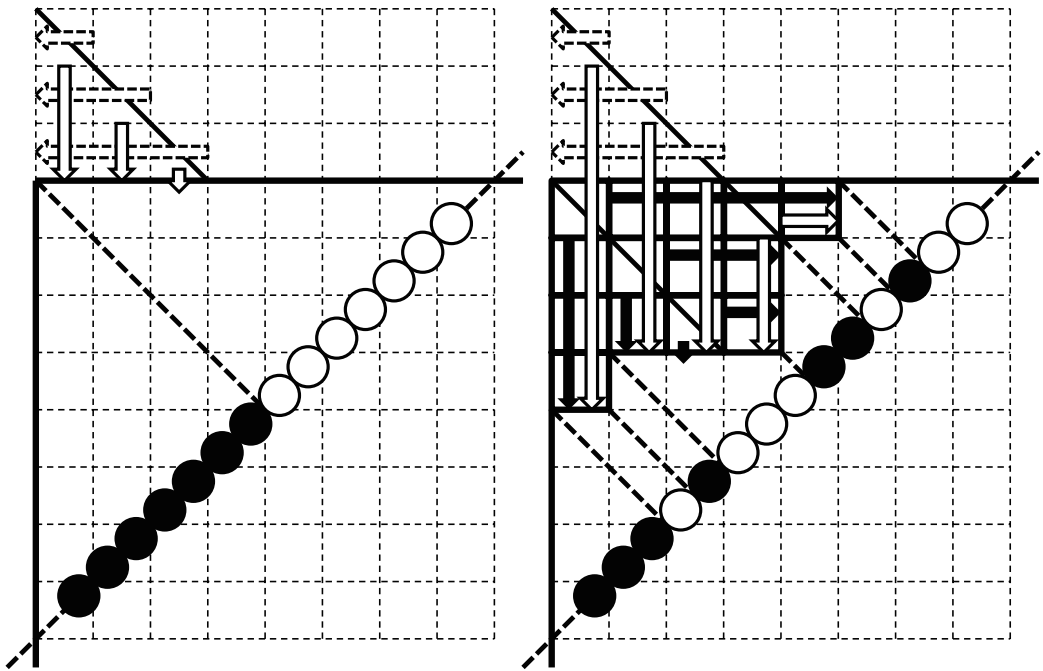}
\caption{Young diagrams, Maya diagrams and Frobenius symbols.
The left figure is the Young diagram in the trivial representation, while the right one is a general Young diagram.
The corresponding Maya diagrams are depicted by drawing a black circle when going vertically and a white circle when going horizontally in tracing the boundary of the Young diagram.
In terms of the Frobenius symbol (counting the arm and leg lengths with the black arrows), the left one is trivial $(|)$ and the right one is $(421|310)$.
In terms of the Frobenius symbol shifted by $M=3$ (counting the lengths with the white solid arrows), the left one is $(|210)$ and the right one is $(1|6432)$.
We often supplement the shifted Frobenius symbol by auxiliary negative arm lengths $(-3,-2,-1)$ (counting with the white dashed arrows), so that there are equal numbers of arm lengths and leg lengths. 
}
\label{maya}
\end{figure}

To state our result, let us start with some preparations.
The Young diagram $Y$ has various expressions,
\begin{align}
Y&=[\lambda_1,\lambda_2,\cdots,\lambda_L]
=[\alpha_1,\alpha_2,\cdots,\alpha_A]^{\rm T}\nonumber\\
&=(a_1,a_2,\cdots,a_r|l_1,l_2,\cdots,l_r)
=(a'_1,a'_2,\cdots,a'_{r'}|l'_1,l'_2,\cdots,l'_{M+r'}).
\end{align}
In the first line, we express it by listing all of positive legs or arms of the Young diagram.
When necessary, we regard $\lambda_{L+1}=\lambda_{L+2}=\cdots=0$.
The second line is called the Frobenius symbol: the first expression is the standard one, counting non-negative horizontal and vertical boxes from the diagonal line, and the next is defined by shifting the diagonal line by $M$,
\begin{align}
a_q=\alpha_q-q,\quad l_p=\lambda_p-p,\quad
a'_q=\alpha_q-q-M,\quad l'_p=\lambda_p-p+M,
\label{armleglength}
\end{align}
with
\begin{align}
r=\max\{s|a_s\ge 0\}=\max\{s|l_s\ge 0\},\quad
r'=\max\{s|a'_s\ge 0\}=\max\{s|l'_s\ge 0\}-M.
\end{align}
We often supplement the shifted Frobenius symbol by auxiliary negative arm lengths 
\begin{align}
b'_q=q-M-1,\quad(1\le q\le M),
\label{auxiliary}
\end{align}
so that there are equal numbers of arm lengths and leg lengths.
See figure \ref{maya}.
With the definition of the Frobenius symbol, the Giambelli identity for the super Schur polynomial $s_{(a_1,a_2,\cdots,a_r|l_1,l_2\cdots,l_r)}$ is expressed as
\begin{align}
s_{(a_1,a_2,\cdots,a_r|l_1,l_2\cdots,l_r)}
=\det\bigl(s_{(a_q|l_p)}\bigr)
_{\begin{subarray}{c}1\le q\le r\\1\le p\le r\end{subarray}}.
\label{GI}
\end{align}
For other quantities labeled by the Young diagram, we propose a criterion for the Giambelli compatibility in the background shift.

\noindent
{\bf Proposition.}
{\it
Let $M$ be a non-negative integer and let $\sigma:{\mathbb Z}\times{\mathbb Z}_{\ge 0}\to{\mathbb C}[z]$ be a set of arbitrary functions labelled by two integers.
Suppose that a quantity $S_Y^M$ is given by
\begin{align}
\frac{S_Y^M}{S_\bullet^0}=\det\begin{pmatrix}
\bigl(\sigma{(b'_q,l'_p)}\bigr)
_{\begin{subarray}{c}1\le q\le M\\1\le p\le M+r'\end{subarray}}\\
\bigl(\sigma{(a'_q,l'_p)}\bigr)
_{\begin{subarray}{c}1\le q\le r'\\1\le p\le M+r'\end{subarray}}
\end{pmatrix},
\label{assumption}
\end{align}
for any Young diagram $Y$, where $b'_q$, $a'_q$ and $l'_p$ are respectively the auxiliary arm length \eqref{auxiliary}, the arm length and the leg length \eqref{armleglength} for the Young diagram $Y$ shifted by $M$.
Then, $S_{(a_1,a_2,\cdots,a_r|l_1,l_2\cdots,l_r)}^M$ normalized by $S_\bullet^M$ satisfies the Giambelli compatibility for general $M$
\begin{align}
\frac{S_{(a_1,a_2,\cdots,a_r|l_1,l_2\cdots,l_r)}^M}{S_\bullet^M}
=\det\biggl(\frac{S_{(a_q|l_p)}^M}{S_\bullet^M}\biggr)
_{\begin{subarray}{c}1\le q\le r\\1\le p\le r\end{subarray}}.
\label{prop}
\end{align}
}Note that, by setting $M=0$, the assumption implies directly the Giambelli compatibility for $M=0$
\begin{align}
\frac{S_{(a_1,a_2,\cdots,a_r|l_1,l_2\cdots,l_r)}^0}{S_\bullet^0}
=\det\biggl(\frac{S_{(a_q|l_p)}^0}{S_\bullet^0}\biggr)
_{\begin{subarray}{c}1\le q\le r\\1\le p\le r\end{subarray}},
\label{zerogi}
\end{align}
if we combine the equation $S_{(a_1,a_2,\cdots,a_r|l_1,l_2\cdots,l_r)}^0/S_\bullet^0=\det_{r\times r}(\sigma{(a_q,l_p)})$ for a general Young diagram $(a_1,a_2,\cdots,a_r|l_1,l_2\cdots,l_r)$ and those for the hook representation, $S^0_{(a_q|l_p)}/S_\bullet^0=\sigma{(a_q,l_p)}$.
The proposition further guarantees the Giambelli compatibility for general $M$. 
As we shall see in the next subsection, this proposition actually helps in studying the Giambelli compatibility for one-point functions of the half-BPS Wilson loop in the ABJM theory.

\subsection{Physical background}\label{phys}

The ABJM theory is the ${\cal N}=6$ superconformal Chern-Simons theory which has gauge group U$(N_1)_k\times$U$(N_2)_{-k}$ (with the indices denoting the Chern-Simons levels) and two pairs of the bifundamental matters \cite{ABJM,HLLLP2,ABJ}.
The theory describes a system of coincident $\min(N_1,N_2)$ M2-branes and $|N_2-N_1|$ fractional M2-branes on a geometry ${\mathbb C}^4/{\mathbb Z}_k$.
After applying the localization theorem \cite{KWY}, the infinite-dimensional path integral in defining the partition function and one-point functions of the half-BPS Wilson loop on $S^3$ reduces to a finite-dimensional multiple integration, called the ABJM matrix model,
\begin{align}
\frac{\llangle s_Y\rrangle(N_1,N_2)}
{(-1)^{\frac{1}{2}N_1(N_1-1)+\frac{1}{2}N_2(N_2-1)}}
=\int\frac{D^{N_1}\mu}{N_1!}\frac{D^{N_2}\nu}{N_2!}
\frac{\prod_{a<b}^{N_1}(2\sinh\frac{\mu_a-\mu_b}{2})^2
\prod_{c<d}^{N_2}(2\sinh\frac{\nu_c-\nu_d}{2})^2}
{\prod_{a=1}^{N_1}\prod_{c=1}^{N_2}(2\cosh\frac{\mu_a-\nu_c}{2})^2}
s_Y(e^\mu,e^\nu),
\end{align}
with the integrations given by
\begin{align}
D\mu_a=\frac{d\mu_a}{2\pi}e^{\frac{ik}{4\pi}\mu_a^2},\quad
D\nu_c=\frac{d\nu_c}{2\pi}e^{-\frac{ik}{4\pi}\nu_c^2}.
\label{integral}
\end{align}
This matrix model is regarded to possess a hidden super gauge group \cite{DT,MPtop} because the integrand is a trigonometric (or hyperbolic) deformation of the U$(N_1|N_2)$ invariant measure and the exponent $\sum_a\mu_a^2-\sum_c\nu_c^2$ of the Fresnel integration \eqref{integral} is a supertrace.
In the integrand, $s_Y(e^\mu,e^\nu)$ is the super Schur polynomial \cite{md}, a character of U$(N_1|N_2)$ labelled by a Young diagram $Y$.
Without loss of generality, we assume that $N_1\le N_2$ and $k>0$.
Otherwise we consider its complex conjugate.
We also define the matrix model in the grand canonical ensemble as \cite{MP,MM}
\begin{align}
\langle s_Y\rangle_M=\sum_{N=0}^\infty z^N
\llangle s_Y\rrangle(N,N+M),
\label{GC}
\end{align}
by introducing fugacity $z$.
Namely, we consider the canonical partition function of $N$ M2-branes and $M$ fractional M2-branes and move to the grand canonical ensemble by transforming $N$ to the dual fugacity $z$.

In \cite{MM} it was shown that the grand canonical matrix model is expressed as
\begin{align}
\frac{\langle s_Y\rangle_M}{\langle 1\rangle_0}
=\det\begin{pmatrix}
\bigl(H_{l'_p,b'_q}\bigr)
_{\begin{subarray}{c}1\le q\le M\\1\le p\le M+r'\end{subarray}}\\
\bigl(\widetilde H_{l'_p,a'_q}\bigr)
_{\begin{subarray}{c}1\le q\le r'\\1\le p\le M+r'\\\end{subarray}}
\end{pmatrix},
\label{HHdet}
\end{align}
with $H_{p,q}$ and $\widetilde H_{p,q}$ defined by
\begin{align}
H_{p,q}=E_p\circ\bigl[1+zQ\circ P\circ\bigr]^{-1}E_q,\quad
\widetilde H_{p,q}
=zE_p\circ\bigl[1+zQ\circ P\circ\bigr]^{-1}Q\circ E_q.
\label{HHdef}
\end{align}
Here $P$, $Q$ and $E_j$ are matrices or vectors with the continuous indices $\mu$ and $\nu$, whose explicit forms are given by
\begin{align}
(P)_{\mu,\nu}=\frac{1}{2\cosh\frac{\mu-\nu}{2}},\quad
(Q)_{\nu,\mu}=\frac{1}{2\cosh\frac{\nu-\mu}{2}},\quad
(E_j)_\nu=e^{(j+\frac{1}{2})\nu}.
\end{align}
In the matrix multiplication $\circ$, we contract the continuous indices by the integrations $D\mu$ or $D\nu$ \eqref{integral}.
The non-negative integers $a'_q$, $l'_p$ are the arm lengths and the leg lengths \eqref{armleglength} appearing in the shifted Frobenius symbol of the Young diagram and the negative integers $b'_q$ are the auxiliary arm lengths \eqref{auxiliary}.
The normalization factor $\langle 1\rangle_0$ is given as
\begin{align}
\langle 1\rangle_0=\Det(1+zQ\circ P\circ),
\label{Det}
\end{align}
with the Fredholm determinant defined by expanding the determinant into traces.

The proof of \eqref{HHdet} is not difficult.
As reviewed in \cite{PTEP}, for the partition function, the basic idea is to use a combination of the Vandermonde determinant and the Cauchy determinant
\begin{align}
\frac{\prod_{a<b}^{N}(x_a-x_b)\prod_{c<d}^{N+M}(y_c-y_d)}
{\prod_{a=1}^{N}\prod_{c=1}^{N+M}(x_a+y_c)}
=(-1)^{NM}\det\begin{pmatrix}
\biggl(\displaystyle\frac{1}{x_a+y_c}\biggr)
_{\begin{subarray}{c}1\le a\le N\\1\le c\le N+M\end{subarray}}\\
\Bigl(y_c^{M-b}\Bigr)
_{\begin{subarray}{c}1\le b\le M\\1\le c\le N+M\end{subarray}}
\end{pmatrix},
\end{align}
to express the integration measure as two determinants of the matrix elements
\begin{align}
&(-1)^{\frac{1}{2}N(N-1)+\frac{1}{2}(N+M)(N+M-1)}
\biggl(\frac{\prod_{a<b}2\sinh\frac{\mu_a-\mu_b}{2}
\prod_{c<d}2\sinh\frac{\nu_c-\nu_d}{2}}
{\prod_{a,c}2\cosh\frac{\mu_a-\nu_c}{2}}\biggr)^2\nonumber\\
&=\det\begin{pmatrix}\bigl((P)_{\mu_a,\nu_c}\bigr)
_{\begin{subarray}{c}1\le a\le N\\1\le c\le N+M\end{subarray}}\\
\bigl((E_{M-b})_{\nu_c}\bigr)
_{\begin{subarray}{c}1\le b\le M\\1\le c\le N+M\end{subarray}}
\end{pmatrix}
\det\begin{pmatrix}\bigl((Q)_{\nu_c,\mu_a}\bigr)
_{\begin{subarray}{c}1\le c\le N+M\\1\le a\le N\end{subarray}}&
\bigl((E_{b-M-1})_{\nu_c}\bigr)
_{\begin{subarray}{c}1\le c\le N+M\\1\le b\le M\end{subarray}}
\end{pmatrix}.
\label{detdet}
\end{align}
Then, the remaining task is to multiply these matrix elements subsequently by contracting the continuous indices $\mu_a,\nu_c$ by $D\mu_a,D\nu_c$ in \eqref{integral}.
As can be easily imagined, the result consists of traces and bilinear terms, which turn out to be summarized as $\langle 1\rangle_0$ \eqref{Det} and $H_{p,q},\widetilde H_{p,q}$ \eqref{HHdef} respectively.
This was done in \cite{MM} by preparing a simple integration formula.
Also, for the one-point functions, we use a determinantal formula \cite{MVdJ} to express the character $s_Y(e^\mu,e^\nu)$ as a ratio of two determinants whose denominator is identical to one of the determinants in \eqref{detdet}.
Then, after the cancellation, we simply replace the denominator determinant in \eqref{detdet} by the numerator determinant, indicating that we can repeat the same contraction as in the partition function.

As a corollary of \eqref{HHdet}, by setting $M=0$ as in \eqref{zerogi}, we find \cite{HHMO}
\begin{align}
\frac{\langle s_{(a_1,a_2,\cdots,a_r|l_1,l_2\cdots,l_r)}\rangle_0}
{\langle 1\rangle_0}
=\det\begin{pmatrix}
\displaystyle
\frac{\langle s_{(a_q|l_p)}\rangle_0}{\langle 1\rangle_0}
\end{pmatrix}
_{\begin{subarray}{c}1\le q\le r\\1\le p\le r\end{subarray}}.
\label{0}
\end{align}
Hence, in \eqref{0} we have seen that we can apply the normalized matrix integration to each character in the Giambelli identity \eqref{GI}.
In order words, the grand canonical one-point functions \eqref{GC} are Giambelli compatible \cite{BOS}.
Our main result in this work is a generalization.

\begin{figure}[!t]
\centering\includegraphics[scale=0.6]{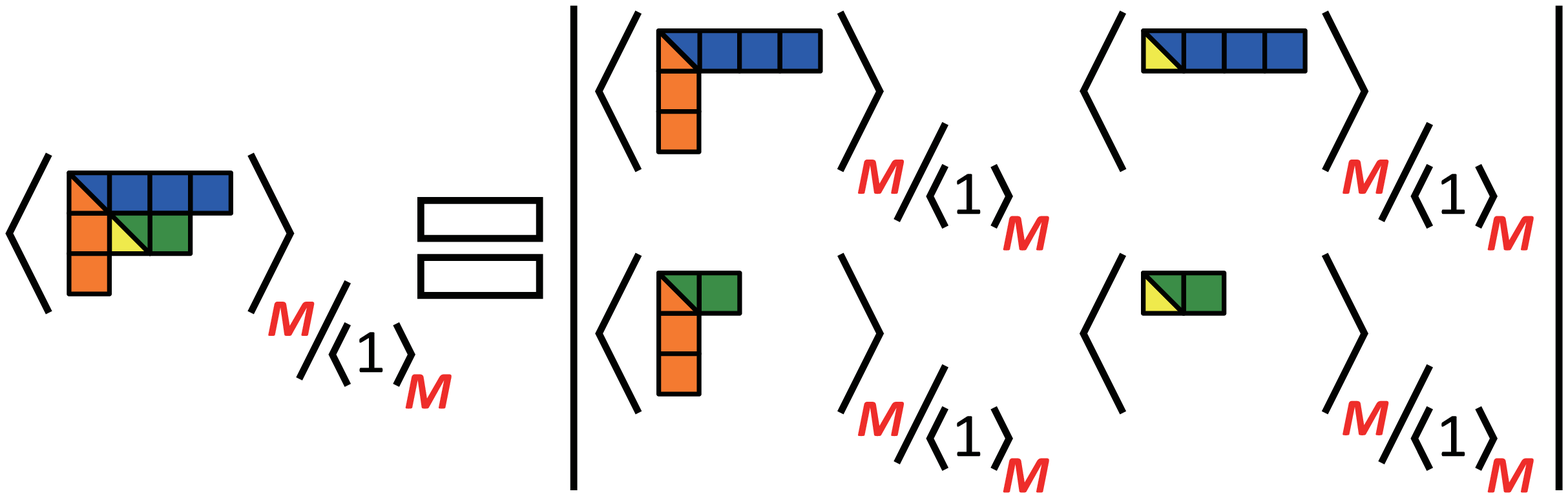}
\caption{A schematic expression for the Giambelli compatibility with a background shift $M$.
}
\label{giambellic}
\end{figure}

\noindent
{\bf Theorem.}
\begin{align}
\frac{\langle s_{(a_1,a_2,\cdots,a_r|l_1,l_2\cdots,l_r)}\rangle_M}
{\langle 1\rangle_M}
=\det\begin{pmatrix}
\displaystyle
\frac{\langle s_{(a_q|l_p)}\rangle_M}{\langle 1\rangle_M}
\end{pmatrix}
_{\begin{subarray}{c}1\le q\le r\\1\le p\le r\end{subarray}}.
\label{Giambelli}
\end{align}
Namely, the Giambelli compatibility is robust under the deformation of the fractional-brane background parametrized by $M$.
See figure \ref{giambellic} for a schematic expression.

It is not difficult to see that, with the expression \eqref{HHdet} at hand, this theorem follows directly from our proposition \eqref{prop} if we set the unknown functions $\sigma$ in \eqref{assumption} to be those appearing in \eqref{HHdet}
\begin{align}
\sigma{(b',l')}=H_{l',b'},\quad
\sigma{(a',l')}=\widetilde H_{l',a'}.
\end{align}
Also, we note that the criterion \eqref{HHdet} is much easier to prove than the original Giambelli compatibility because this formula comes simply from a combinatorics in the contractions, as we review in \eqref{detdet}.

In addition, our proposition seems applicable to more general situations.
In fact, as long as a determinantal expression corresponding to \eqref{HHdet} is valid, we can apply our proposition directly to other Chern-Simons matrix models such as the orthosymplectic matrix model \cite{HMO1,MS1,Ho2,Ok2,MS2,MN5}, coming from the ${\cal N}=5$ orthosymplectic theory \cite{HLLLP2,ABJ}, various ${\cal N}=4$ models \cite{HoMo,MN1,MN2,MN3,HHO}, coming from the ${\cal N}=4$ Chern-Simons theories \cite{IK4}, or even more general $\widehat A$ \cite{MP} or $\widehat D$ \cite{ADF,MN4} matrix models, coming from such quiver Chern-Simons theories.

\section{Proof}\label{pf}

We prove the proposition in this section.
Under the assumption in the proposition, we shall prove
\begin{align}
S_{(a_1,a_2,\cdots,a_r|l_1,l_2,\cdots,l_r)}^M(S_\bullet^M)^{r-1}
=\det(S_{(a_q|l_p)}^M)
_{\begin{subarray}{c}1\le q\le r\\1\le p\le r\end{subarray}}.
\label{det}
\end{align}
We start with our proof in a comparatively simple situation and then turn to the general case.
Hereafter, for simplicity, we abbreviate the determinant symbol $\det(\cdots)$ by $|\cdots|$.

\subsection{$r'=0$ case}

Let us first consider the case when $M$ is large enough so that the shifted diagonal line does not have any overlaps with the Young diagram.
In other words, $r'=0$ and the block $\bigl(\sigma{(a'_q,l'_p)}\bigr)$ does not appear in \eqref{assumption}.
In this case the (supplemented) arm lengths are always $(-M,-(M-1),\cdots,-1)$ and we shall specify the leg lengths $(l'_1,l'_2,\cdots,l'_M)$ only.
This set of the leg lengths has $M-r$ overlaps with the trivial ones $(M-1,M-2,\cdots,0)$ for $S^M_\bullet$ since $l'_r\ge M> l'_{r+1}$.
We shall split the set of leg lengths into the non-overlapping subset and the overlapping subset
\begin{align}
(l'_1,l'_2,\cdots,l'_M)
&=(l'_1,l'_2,\cdots,l'_r)\sqcup(l'_{r+1},l'_{r+2},\cdots,l'_M),
\nonumber\\
(M-1,M-2,\cdots,0)
&=(m_1,m_2,\cdots,m_r)\sqcup(l'_{r+1},l'_{r+2},\cdots,l'_M),
\end{align}
with $m_q>m_{q+1}$.
Then, it turns out that
\begin{align}
m_q=M-1-a_{r+1-q},\quad(1\le q\le r).
\label{m}
\end{align}
This is true because of the following reason.
When $m_q$ with $0\le m_q\le M-1$ is missing in $(l'_{r+1},l'_{r+2},\cdots,l'_M)$, there are no horizontal segments in the boundary of the Young diagram, which are distant from the shifted diagonal line by $m_q$.
For the boundary of the Young diagram to be connected, we need vertical segments which are distant by $m_q$.
This means that $\{m_q\}_{1\le q\le r}$ is the set of distances of the vertical segments from the shifted diagonal line, since both have $r$ elements.
In other words, $\{M-1-m_q\}_{1\le q\le r}$ is the set of distances of the vertical segments from the original diagonal line, which are nothing but the arm lengths.
Intuitively, when a leg length in $(M-1,M-2,\cdots,0)$ is missing in $(l'_{r+1},l'_{r+2},\cdots,l'_M)$, we have a vertical jump and this jump reproduces an arm length. (See subsection \ref{ex} and figure \ref{example} for a pictorial explanation.)
In this way, we can effectively transform the information of the leg lengths to the arm lengths.

If we rewrite $S_{(a_q|l_p)}^M$ for the hook representation appearing in the determinant in \eqref{det} in terms of the shifted Frobenius symbol, we find
\begin{align}
\frac{S_{(a_q|l_p)}^M}{S_\bullet^0}
=|\z_{l'_p}\z_{M-1}\z_{M-2}\cdots\check\z_{M-1-a_q}\cdots\z_{0}|,
\end{align}
where we denote $(\z_{l'_p})_q=\sigma{(b'_q,l'_p)}$ and the check in $\check\z$ means the removal of the corresponding column.
The reason for the appearance of the arm length $a_q$ is the same as \eqref{m}.
Then, the identity we want to prove \eqref{det} becomes
\begin{align}
&|\z_{l'_1}\z_{l'_2}\cdots\z_{l'_M}|
|\z_{M-1}\z_{M-2}\cdots\z_{0}|^{r-1}
=\biggl|\bigl(|\z_{l'_p}\z_{M-1}\z_{M-2}\cdots
\check\z_{M-1-a_q}\cdots\z_0|\bigr)
_{\begin{subarray}{c}1\le q\le r\\1\le p\le r\end{subarray}}\biggr|,
\label{r0}
\end{align}
We shall prove \eqref{r0} by the following lemma, which is proved in the next subsection.

\noindent
{\bf Lemma.}
{\it The determinant formula
\begin{align}
|\x_1\x_2\cdots\x_rA|
|\y_1\y_2\cdots\y_rA|^{r-1}
&=\left|\begin{pmatrix}
|\x_i\y_1\y_2\cdots\check\y_{r+1-j}\cdots\y_rA|
\end{pmatrix}
_{\begin{subarray}{c}1\le j\le r\\1\le i\le r\end{subarray}}
\right|\nonumber\\
&=\left|\begin{matrix}
|\x_1\y_1\y_2\cdots\y_{r-1}A|&\cdots
&|\x_r\y_1\y_2\cdots\y_{r-1}A|\\
\vdots&&\vdots\\
|\x_1\y_2\y_3\cdots\y_rA|&\cdots
&|\x_r\y_2\y_3\cdots\y_rA|
\end{matrix}\right|,
\label{lemma}
\end{align}
holds where $\x$ and $\y$ are column vectors and $A$ is a collection of vectors.}

If we apply the lemma \eqref{lemma} with
\begin{align}
\x_p=\z_{l'_p},\quad\y_q=\z_{m_q},\quad
A=(\z_{l'_{r+1}}\z_{l'_{r+2}}\cdots\z_{l'_M}),
\end{align}
we almost reproduce \eqref{r0} correctly (except for the difference in the order of various vectors $\z$), since on the right-hand side we remove the vector with index
\begin{align}
{m_{r+1-q}}={M-1-a_{r+1-(r+1-q)}}={M-1-a_q},
\end{align}
from the set $(m_1,m_2,\cdots,m_r)\sqcup(l'_{r+1},l'_{r+2},\cdots,l'_M)=(M-1,\cdots,0)$.
When changing the indices into the correct decreasing order we have to move $(m_1,m_2,\cdots,m_r)\sqcup(l'_{r+1},l'_{r+2},\cdots,l'_M)$ to
$(M-1,M-2,\cdots,0)$.
If we assign a sign factor $(-1)^{s_q}$ to each $m_q$ counting the numbers of transpositions appearing in the permutation with $(l'_{r+1},\cdots,l'_M)$, we encounter $(-1)^{(r-1)\sum_{q=1}^rs_q}$ on the left-hand side of \eqref{r0}, while $(-1)^{\sum_{q=1}^r\sum_{q'\ne q}s_{q'}}$ on the right-hand side.
Apparently, these two factors cancel each other and we obtain \eqref{det} for the case $r'=0$ finally.

\subsection{A determinant formula}\label{detid}

We shall prove the determinant formula \eqref{lemma} in this subsection.
We can prove it by induction with respect to the number of $\x$ and $\y$.
(The size of the columns $\x$, $\y$ or $A$ can be arbitrary.)

For $r=2$, the identity follows from the Laplace expansion
\begin{align}
\left|\begin{matrix}
\x_1\x_2\y_1\y_2AA\\
\x_1\x_2\y_1\y_2AA\\
\end{matrix}\right|
=|\x_1\x_2A||\y_1\y_2A|-|\x_1\y_1A||\x_2\y_2A|+|\x_1\y_2A||\x_2\y_1A|,
\end{align}
which is vanishing trivially.

Assuming this is true for $r=n$, for $r=n+1$ we first Laplace-expand the right-hand side along the first row and apply the assumption of the induction by regarding $\y_{n+1}A$ as a new collection of vectors,
\begin{align}
&\biggl|\bigl(|\x_i\y_1\y_2\cdots\check\y_{n+2-j}\cdots\y_{n+1}A|\bigr)
_{\begin{subarray}{c}1\le j\le n+1\\1\le i\le n+1\end{subarray}}\biggr|
\nonumber\\
&=\sum_{i=1}^{n+1}(-1)^{i-1}|\x_i\y_1\y_2\cdots\y_nA|
\biggl|\bigl(|\x_{k}\y_1\cdots\check\y_{n+1-j}\cdots\y_n\y_{n+1}A|\bigr)
_{\begin{subarray}{c}1\le j\le n\\
1\le k\le i-1,i+1\le k\le n+1\end{subarray}}
\biggr|
\nonumber\\
&=\sum_{i=1}^{n+1}(-1)^{i-1}|\x_i\y_1\y_2\cdots\y_nA|
|\x_1\x_2\cdots\check\x_i\cdots\x_{n+1}\y_{n+1}A|
|\y_1\y_2\cdots\y_n\y_{n+1}A|^{n-1}
\nonumber\\
&=\Biggl[-\left|\begin{matrix}
\x_1\x_2\cdots\x_{n+1}&A&0\;\;0\;\cdots\;0\:&\y_{n+1}&O\\
\x_1\x_2\cdots\x_{n+1}&A&\y_1\y_2\cdots\y_n&\y_{n+1}&A
\end{matrix}\right|
+|\x_1\x_2\cdots\x_{n+1}A|
|\y_1\y_2\cdots\y_n\y_{n+1}A|\Biggr]
\nonumber\\
&\quad\times
|\y_1\y_2\cdots\y_n\y_{n+1}A|^{n-1}\nonumber\\
&=|\x_1\x_2\cdots\x_{n+1}A||\y_1\y_2\cdots\y_n\y_{n+1}A|^{n}.
\end{align}
In the third equality we introduce a vanishing determinant of double size whose Laplace expansion gives only one additional term compared with our original expression.
The determinant of double size is vanishing because of the following reason.
After changing the columns it takes the form
\begin{align}
\left|\begin{matrix}X_{(n+a+1)\times(n+a+2)}&O_{(n+a+1)\times(n+a)}\\
X_{(n+a+1)\times(n+a+2)}&Y_{(n+a+1)\times(n+a)}\end{matrix}\right|
=\sum_{i=1}^{n+a+1}(-1)^{i-1}
\left|\begin{matrix}\bigl(X\bigr)_{(n+a+1)\times(n+a+2)}\\
\bigl(X_i\bigr)_{1\times(n+a+2)}\end{matrix}\right|
\Bigl|\bigl(Y_{\check i}\bigr)_{(n+a)\times(n+a)}\Bigr|,
\end{align}
where $a$ is the number of the vectors in $A$, $X_i$ is the $i$-th row of the matrix $X$, while $Y_{\check i}$ is the matrix obtained from $Y$ after removing the $i$-th row.
The right-hand side is vanishing because the first determinant contains two identical rows.
This completes our induction.

\subsection{$r'\ne 0$ case}

In this case apparently the determinants coming from $S_{(a_1,a_2,\cdots,a_r|l_1,l_2\cdots,l_r)}^M$ and $S_\bullet^M$ have different sizes and we cannot apply the determinant formula \eqref{lemma} directly.
For this reason, we shall increase the sizes of the matrices in the determinants artificially so that we can apply \eqref{lemma} and then remove the extra components afterwards.
Namely, we shall fix the matrix size $M+r'$ defined from $S_{(a_1,a_2,\cdots,a_r|l_1,l_2\cdots,l_r)}^M$ and fill the empty columns in other determinants with minus leg lengths so that the whole matrix size does not change.
More concretely, we shall fix the set of shifted arm lengths to be that for the original Young diagram $(a_1,\cdots,a_r|l_1,\cdots,l_r)$,
\begin{align}
(-M,-(M-1),\cdots,-1,a'_1,a'_2,\cdots,a'_{r'}),
\end{align}
and increase the number of the leg lengths for $S_\bullet^M$ formally into
\begin{align}
(M-1,M-2,\cdots,0,-1,-2,\cdots,-r'),
\label{leglength}
\end{align}
with the minus leg lengths.
Again, since $l'_r\ge M>l'_{r+1}$, the set of leg lengths has $M+r'-r$ overlaps with \eqref{leglength}.
Hence, as previously, we can separate two sets as
\begin{align}
(l'_1,l'_2,\cdots,l'_M,l'_{M+1},l'_{M+2},\cdots,l'_{M+r'})
&=(l'_1,l'_2,\cdots,l'_r)\sqcup(l'_{r+1},l'_{r+2},\cdots,l'_{M+r'}),
\nonumber\\
(M-1,\cdots,1,0,-1,-2,\cdots,-r')&=
(m_1,\cdots,m_{r-r'})\sqcup(-1,\cdots,-r')
\sqcup(l'_{r+1},\cdots,l'_{M+r'}).
\label{separate}
\end{align}
Then we find that
\begin{align}
m_q&=M-1-a_{r+1-q},\quad (1\le q\le r-r'),
\end{align}
for the same reason as \eqref{m}, $\{M-1-m_q\}_{1\le q\le r-r'}=\{a_q\}_{r'+1\le q\le r}$.
Using \eqref{lemma} for the leg lengths with the overlapping ones separated \eqref{separate} and moving to the correct decreasing order as in the case of $r'=0$, we arrive at an identity
\begin{align}
&|\z_{l'_1}\z_{l'_2}\cdots\z_{l'_{M+r'}}|
|\z_{M-1}\z_{M-2}\cdots\z_0\z_{-1}\z_{-2}\cdots\z_{-r'}|^{r-1}
\nonumber\\
&=\left|\begin{matrix}
\bigl(|\z_{l'_p}\z_{M-1}\cdots\z_0
\z_{-1}\cdots\check\z_{-r'-1+q}\cdots\z_{-r'}|\bigr)
_{\begin{subarray}{c}1\le q\le r'\\1\le p\le r\end{subarray}}\\
\bigl(|\z_{l'_p}\z_{M-1}\cdots\check\z_{M-1-a_{q}}\cdots\z_0
\z_{-1}\cdots\z_{-r'}|\bigr)
_{\begin{subarray}{c}r'+1\le q\le r\\1\le p\le r\end{subarray}}
\end{matrix}\right|.
\end{align}
Here for the lower block we remove the vector with the leg length
\begin{align}
m_{r-r'+1-q}=M-1-a_{r'+q},\quad(1\le q\le r-r'),
\end{align}
from $(m_1,\cdots,m_{r-r'})\sqcup(l'_{r+1},\cdots,l'_{M+r'})=(M-1,\cdots,0)$.
Though in the permutations we need to consider extra sign factors $(-1)^{M+r'-r}$ assigned to each negative leg length, the cancellation between both sides happens in the same way.

Since this is a general identity valid for any component, we are free to reduce the matrix size by setting
\begin{align}
(\z_{-p})_q=\begin{cases}1,\quad&q=M+p,\\0,\quad&q\ne M+p,\end{cases}
\end{align}
for negative leg lengths $1\le p\le r'$, while keeping 
\begin{align}
(\z_{l'_p})_q=\begin{cases}\sigma{(b'_q,l'_p)},\quad&1\le q\le M,\\
\sigma{(a'_{q-M},l'_p)},\quad&M+1\le q\le M+r',\end{cases}
\end{align}
for $1\le p\le M$.
Then, each component reduces exactly to the hook diagram appearing in the determinant.
The component in the lower block reduces to $|\z_{l'_p}\z_{M-1}\z_{M-2}\cdots\check\z_{M-1-a_{q}}\cdots\z_0|$ with arm lengths $(-M,-(M-1),\cdots,-1)$, which is nothing but $S_{(a_q|l_p)}^M$.
The component in the upper block reduces to $(-1)^{r'-q}|\z_{l'_p}\z_{M-1}\cdots\z_0|$ with the arm lengths $(-M,-(M-1),\cdots,-1,a'_{r'+1-q})$ which is $(-1)^{r'-q}S_{(a_{r'+1-q}|l_p)}^M$,
\begin{align}
S_{(a_1,\cdots,a_r|l_1,\cdots,l_r)}^M
(S_\bullet^M)^{r-1}
=\left|\begin{matrix}
\bigl((-1)^{r'-q}S_{(a_{r'+1-q}|l_p)}^M\bigr)
_{\begin{subarray}{c}1\le q\le r'\\1\le p\le r\end{subarray}}\\
\bigl(S_{(a_{q}|l_p)}^M\bigr)
_{\begin{subarray}{c}r'+1\le q\le r\\1\le p\le r\end{subarray}}
\end{matrix}\right|.
\end{align}
This result is almost \eqref{det} except for the extra signs $(-1)^{r'-q}$ and the reversing of the arm lengths in $S_{(a_{r'+1-q}|l_p)}^M$.
Both of these corrections give a sign factor $(-1)^{\sum_{q=1}^{r'}(r'-q)}=(-1)^{\sum_{q=0}^{r'-1}q}$ and cancel each other.
Finally we obtain \eqref{det} for $r'\ne 0$ correctly.

\subsection{An example}\label{ex}

\begin{figure}[!ht]
\centering\includegraphics[scale=0.6,angle=-90]{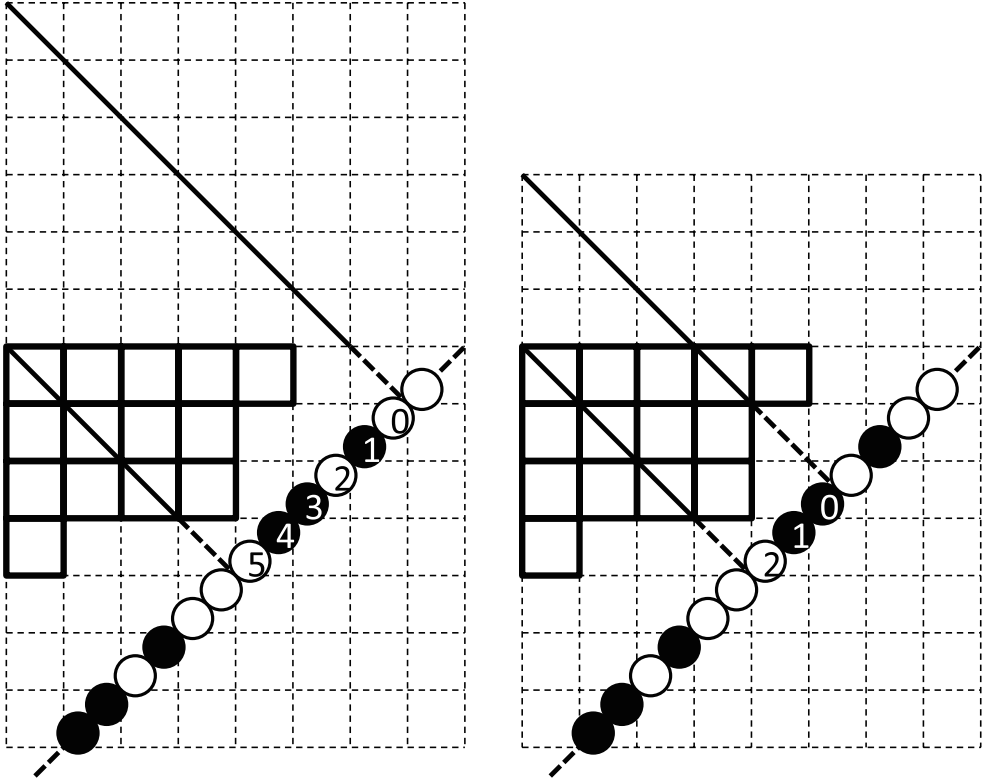}
\caption{An example of the same Young diagram $(421|310)$ with different backgrounds $M=6$ (left) and $M=3$ (right).
Black and white circles between two diagonal lines denote the splitting $(543210)=(431)\sqcup(520)$ for $M=6$ and $(210)=(10)\sqcup(2)$ for $M=3$.}
\label{example}
\end{figure}

In this subsection we shall present an example of our proof.
We shall consider the previous Young diagram $(421|310)$ in the Frobenius symbol.
See figure \ref{example}.

For the case of $M=6$ which is large enough so that $r'=0$, we start with an identity proved in subsection \ref{detid},
\begin{align}
|976520||431520|^2=\left|\begin{matrix}
|943520|&|743520|&|643520|\\
|941520|&|741520|&|641520|\\
|931520|&|731520|&|631520|
\end{matrix}\right|,
\end{align}
where we have abbreviated $\z_i$ simply as $i$ and denoted negative integers by bars.
This relation is reshuffled into
\begin{align}
|976520||543210|^2=\left|\begin{matrix}
|954320|&|754320|&|654320|\\
|954210|&|754210|&|654210|\\
|953210|&|753210|&|653210|
\end{matrix}\right|,
\end{align}
without introducing any signs.

Next, we consider the case of $M=3$.
In this case, the arm lengths are $(-3,-2,-1,1)$, though for the leg lengths, since $r'=1$, we need to introduce one negative leg length as in $(2,1,0,-1)$ for $S_\bullet^3$.
Then, after the reshuffling explained previously, we have
\begin{align}
|6432||210\bar 1|^2
=\left|\begin{matrix}
|6210|&|4210|&|3210|\\
|621\bar 1|&|421\bar 1|&|321\bar 1|\\
|620\bar 1|&|420\bar 1|&|320\bar 1|
\end{matrix}\right|.
\end{align}
Finally we follow the previous rule of reduction to obtain
\begin{align}
|6432||210|^2
=\left|\begin{matrix}
|6210|&|4210|&|3210|\\
|621|&|421|&|321|\\
|620|&|420|&|320|
\end{matrix}\right|,
\end{align}
where the arm lengths are $(-3,-2,-1,1)$ for the component with 4 vectors, while $(-3,-2,-1)$ for the component with only 3 vectors.

\section{Discussion}\label{conclusion}

In this paper, we have proved the Giambelli compatibility for general $M$ in the ABJM matrix model.
The physical origin of this matrix model is one-point functions of the half-BPS Wilson loops on $S^3$ in the ${\cal N}=6$ supersymmetric Chern-Simons theory \cite{ABJM,HLLLP2,ABJ} which has gauge group U$(N)_k\times$U$(N+M)_{-k}$ and two pairs of bifundamental matters with the subscripts $(k,-k)$ denoting the Chern-Simons levels.
This theory describes the worldvolume of $N$ M2-branes and $M$ fractional M2-branes on ${\mathbb C}^4/{\mathbb Z}_k$.  
After applying the localization theorem \cite{KWY}, the infinite dimensional path integral reduces to the finite dimensional matrix integral.
Hence, our main claim in this paper is that the Giambelli compatibility holds independent of the background parametrized by the number of the fractional M2-branes $M$.
In short, the Giambelli compatibility is background independent.
See also \cite{MM,M}.

Our proof is general in the sense that we only rely on the determinantal expression proved in \cite{MM}.
Hence, as long as a corresponding determinantal expression is valid, our proposition is applicable to many other Chern-Simons matrix models, such as the orthosymplectic matrix model \cite{HMO1,MS1,Ho2,Ok2,MS2,MN5} and so on.
Probably, this means that the Giambelli identity reflects only the symmetry of the system and is very robust independent of the fractional-brane backgrounds labeled by $M$, the gauge groups or the quivers, without referring to the exact large $N$ expansions \cite{DMP1,FHM,MP,HMO2,HMO3,HMMO,HHMO}.

Our starting point \eqref{HHdet} views the Wilson loop on a fractional brane background $M$ from a trivial background \cite{M}, while the Giambelli compatibility \eqref{Giambelli} views the same Wilson loop directly from the fractional-brane background $M$.
It is possible to prove a similar identity by viewing the same Wilson loop from other fractional-brane backgrounds.
We would like to pursue these directions more extensively.

\subsection*{Note added}

After we have finished the proof and prepare the manuscript, we are informed by Kazumi Okuyama of their paper \cite{HO} where the Giambelli identity was checked numerically for several cases.

\subsection*{Acknowledgements}

We are grateful to Kazumi Okuyama, Takeshi Oota and especially Sho Matsumoto and Masato Okado for valuable discussions.
The work of Sa.Mo.\ is supported by JSPS Grant-in-Aid for Scientific
Research (C) \# 26400245.

\end{document}